\documentstyle[12pt,a4wide]{article}
\let\ssection=\section
\renewcommand{\section}{\setcounter{equation}{0}\ssection}

\newcommand{\be}{\begin{enumerate}}
\newcommand{\ee}{\end{enumerate}}
\newcommand{\bi}{\begin{itemize}}
\newcommand{\ei}{\end{itemize}}
\newcommand{\beq}{\begin{equation}}
\newcommand{\eeq}{\end{equation}}
\newcommand{\beqa}{\begin{eqnarray}}
\newcommand{\eeqa}{\end{eqnarray}}

\renewcommand{\a}{\alpha}                   
\renewcommand{\b}{\beta}                    
\renewcommand{\d}{\delta}               
\newcommand{\g}{\gamma}
\newcommand{\k}{\kappa}
\renewcommand{\l}{\lambda}

\newcommand{\bra}[1]{\langle #1 |}

\newcommand{\ket}[1]{| #1 \rangle}

\newcommand{\tr}{{\rm\, tr}}

\newcommand\mathC{\mkern1mu\raise2.2pt\hbox{$\scriptscriptstyle|$}
                {\mkern-7mu\rm C}}
\newcommand{\mathR}{{\rm I\! R}}                

\newcommand{\C}{{\tilde{C}}}
\newcommand{\D}{{\cal D}}

\newcommand{\PV}{{\cal P}({\cal V})}
\newcommand{\UP}{{\cal UP}}
\renewcommand{\H}{{\cal H}}
\newcommand{\V}{{\cal V}}

\newcommand{\h}[1]{({#1}_{t_1},{#1}_{t_2},\ldots,{#1}_{t_n})}

\begin{document}

\begin{titlepage}
\hspace{8,5truecm} Imperial/TP/94--95/22 \\
\hspace*{10,3truecm} gr-qc/9502036
\smallskip\\

\begin{center}
   {\large\bf  Completeness of   Decoherence Functionals}

\end{center}
\smallskip
\begin{center}
                Stephan Schreckenberg\footnote{email: stschr@ic.ac.uk}\\[0.5cm]
                Blackett Laboratory\\
                Imperial College\\
                South Kensington\\
                London SW7 2BZ\\[0.2cm]
\end{center}

\begin{center} February 1995\end{center}
\vfill

\begin{abstract}
The basic ingredients of the `consistent histories' approach to
a  generalized quantum theory are `histories'
and      decoherence functionals. The main aim of this program is to find
and to study the behaviour of
consistent sets associated with a
particular decoherence functional $d$. In its recent formulation by Isham
\cite{I94a} it is natural to
identify the space $\UP$ of propositions about histories with an
orthoalgebra or lattice.
When $\UP$ is
given by the
lattice of projectors $\PV$ in some Hilbert space
$\V$, consistent sets correspond to certain partitions of the unit
operator in $\V$ into mutually orthogonal projectors
$\{\a_1,\a_2,\ldots\}$, such that the function $d(\a,\a)$ is a
probability distribution on the boolean algebra generated by
$\{\a_1,\a_2,\ldots\}$.
Using the classification theorem for decoherence functionals proven in
\cite{ILS94c} we show that in the case where $\V$ is some
separable Hilbert space there exists for
each partition of
the unit operator into a set of mutually orthogonal projectors, and for
any probability distribution $p(\a)$ on the corresponding boolean
algebra, decoherence functionals $d$ with respect to which this set is
consistent and which are such that for the probability functions
$d(\a,\a)=p(\a)$ holds.\\

 \end{abstract}

\end{titlepage}

\section{INTRODUCTION}
\label{Sec:intro}

Over the past three years the `consistent histories' approach to quantum
theory, initiated by the work of Griffiths \cite{Gri84}, Omn\`{e}s
\cite{Omn88a}, Gell-Mann and
Hartle \cite{GH90a}, has been subject of intense research. The motivations for
these
activities are diverse and include hopes that the new
conceptual framework advocated will be
general enough to resolve issues connected with the measurement
problem and the
quantum theory of closed systems, as well as giving new insights into
the problem of constructing a quantum theory of gravity.

    The two basic ingredients in this scheme are a `space of
histories' $\UP$ and a `space of decoherence functionals' $\D$, whose
elements are complex-valued functions on pairs of histories. \\
Recall how these spaces arise in non-relativistic quantum
mechanics.
Let us suppose we are given a Hilbert space $\H$, a Hamiltonian
operator $H$ and a density matrix $\rho_{t_0}$ decribing the
state at some initial time $t_0$. Then the joint probability of
finding all the properties  $\a:=\h{\a}$ with $t_1<t_2<\cdots<t_n$,
given by the Schr\"odinger picture projection operators $\{\a_{t_i}\}$,  in a
time-ordered sequence of measurements is given by:
\beq
        d_{(H,\rho)}(\a,\a)=
            \tr_\H(\C_\a^\dagger\rho_{t_0} \C_\a),      \label{Prob:a1-an}
\eeq
where the `class' operator $\C_\a$ is defined to be
\beq
\C_\a:=\a_{t_1}(t_1)\a_{t_2}(t_2)\cdots\a_{t_n}(t_n)
\eeq
with
$\{\a_{t_i}(t_i):=
e^{\frac{i}{\hbar}H(t_i-t_0)}\a_{t_i}e^{-\frac{i}{\hbar}H(t_i-t_0)}\}$
being the associated Heisenberg picture operators. This set
$\a:=\h{\a}$ is called a homogeneous history.\\

In the consistent histories approach one assumes that
the assignment of probabilities to histories is still meaningful but
{\em only} for certain subsets of the space of histories. These subsets
are selected through the requirement that the central object, a
complex-valued {\em decoherence functional} $d(\a,\b)\in\D$, like for
example
 \beq
        d_{(H,\rho)}(\a,\b)=
            \tr_\H(\C_\a^\dagger\rho_{t_0} \C_\b),\label{df}
\eeq
satisfies the so called {\em consistency conditions} when evaluated on
pairs $(\a,\b)$ of elements of those subsets. The virtue is that one
can talk about
the elements of consistent sets  as posessing definite properties.
Notice that, as suggested by the notation, the initial state,
$\rho_{t_0}$, and the dynamical
structure, i.e.\ $H$, are coded into the decoherence functional
$d\in\D$. This analysis in terms of a `passive' set of histories and
an  `active' set of decoherence functionals allows us to make precise
statements about both spaces. \\

   In order to give a proper meaning to the expressions just introduced,
as well as to the heuristic operations of taking disjoint sums, negation and
coarse-graining on the space of histories for the general theory, as
they had been introduced
by Gell-Mann and Hartle, it was first suggested by Isham in \cite{I94a}
that the set of histories $\UP$, more precisely the set of
propositions about `histories' or `universes',  can be described by an
orthoalgebra.\\
The argument for proposing this structure is that the minimal
mathematical object on which such relations can be defined seems to be
an orthoalgebra
whose algebraic operations $(\oplus,\neg,\le)$ correspond to the
aforementioned operations and which has elements ${\bf 0,1}\in\UP$
such that ${\bf 0}\le\a\le{\bf 1}$ for all $\a\in\UP$.\\ Note that the word
`history' or `universe' is just a label for elements of $\UP$; it does
{\em not}
necessarily imply any temporal properties.

An example where such an algebra arises is provided by standard
quantum mechanics when formulated as a history theory. In this case
the set of all
history-propositions $\UP$  can indeed be
identified with the lattice of projectors on a certain Hilbert space
(\cite{I94a,IL94b} and section IV). \\

In terms of the algebraic structure on the space
$\UP$ the defining properties for decoherence functionals are as
follows.

Any decoherence functional $d:\UP\times\UP\rightarrow\mathC$ has to
satisfy the following conditions:
    \be
    \item {\em Hermiticity\/}: $d(\a,\b)=d(\b,\a)^*$ for all
          $\a,\b$.
    \item {\em Positivity\/}: $d(\a,\a)\ge0$ for all $\a$.
    \item {\em Additivity\/}: $d(\a\oplus\b,\g)=d(\a,\g)+d(\b,\g)$ for
          all $\g$, whenever $\a$ and $\b$ are disjoint. If
          appropriate, this can be extended to countable sums.
    \item {\em Normalisation\/}: $d(1,1)=1$.
    \ee

We emphazise that from these requirements
alone it does {\em not}
necessarily follow that in the case of standard quantum mechanics these
functionals must be of the form (\ref{df}), even
if one adds a final density matrix or replaces the unitary evolution by a
non-unitary one \cite{IL94b}.\\

In the `consistent histories' programme one is particularly interested
in boolean subalgebras $\cal A$ of $\UP$, which contain the unit
element and which are associated with a particular decoherence
functional $d$ in such a way that the function $d(\a,\a)$ is a probability
distribution on $\cal A$. To find such algebras one imposes `consistency
conditions' on the values of the decoherence functional $d$ on
pairs of history
propositions which can serve as generators of boolean
subalgebras of $\UP$. If an algebra $\cal A$ is found by this
procedure its generators are often referred to  as a
`consistent set of history propositions with respect to $d$'.\\

\noindent
Therefore, a consistent set obtained in such a way carries two pieces of
information:
\be
\item It is a {\em consistent} set; it has been obtained by fulfilling
the consistency conditions associated with a particular decoherence
functional $d\in\D$.
\item It carries a probability function $p_d(\a):=d(\a,\a)$.
\ee
\noindent

The aim of the present paper is to investigate whether or not this
information  in turn is
sufficient to determine the decoherence
functional $d$ from which this set arose. More specifically, we ask the
following questions:
\bi
\item For which boolean subalgebra $\cal A$ of $\UP$ and which probability
distributions $p(\a)$ on $\cal A$ does there exist a decoherence
functional $d\in\D$ such
that this pair can be obtained from a consistent set with respect to $d$?
\item If such a decoherence functional $d$ exists, is it uniquely
associated with the pair $(\cal A$$, p(\a))$?
\ei
\noindent
Given any separable Hilbert space $\V$, its lattice of projection
operators $\PV$ forms an orthoalgebra and can thus serve as an example
for the space $\UP$. It is therefore of considerable interest to
answer these questions for  this particular model of the
proposed framework. \\

It turns out that in this case for {\em any} such  pair of a boolean
algebra and a
  probability distribution there exist in fact {\em many} decoherence
functionals which serve the purpose.
This result arises as a corollary of the
classification theorem for decoherence functionals proven in
\cite{ILS94c}. The value of this rather technical statement stems from
its implications for the formalism to be developed.\\

The paper is organized as follows. In section II we provide some
basic definitions and results to prepare the main result which will be
proven in section III. In section IV we discuss some
of its implications and finish this paper
with conclusions drawn in section V.

\section{DEFINITIONS AND BASIC FACTS}
\label{Sec BF}

We are now going to define the notion of a consistent set of history
propositions. \\

{\bf Definition:}
We call a set of history propositions $I_n:=\{\a_1,\a_2,\ldots,\a_n\}$
a partition of unity, if all the $\{\a_i\}$ are mutually orthogonal
and add up to the unit in $\UP$, that is  if ${\bf
1}=\a_1\oplus\a_2\oplus\cdots\oplus\a_n$.\\

A set of history propositions that has the property of being a partition
of unity is often
referred to as being
{\em exclusive}, that is its elements are mutually orthogonal, and
{\em exhaustive}, that is $\oplus_{i=1}^n\a_i={\bf 1}$.\\

In case when $\UP$ is the lattice of projectors of an
infinite-dimensional Hilbert space the equality has to be substituted
by the
requirement of weak convergence of a disjoint union of a finite or countable
number of projectors to the unit operator.\\

{\bf Definition:} Fix a decoherence functional $d\in\D$. If there
exists a partition of unity $I_n$ in $\UP$, such that for all $\a_i\in
I_n$ the consistency conditions given below hold, $I_n$ is
called a {\em consistent set of history propositions with respect to the
decoherence functional $d\in\D$}.\\
The conditions are:
\beq
d(\a_i,\a_j)=0 \quad\mbox{for all}\quad i\neq j ;
i,j\in\{1,2,\ldots,n\}. \label{CS}
\eeq

If there exists for a decoherence functional $d\in\D$ a
consistent set $I_n$, then the properties of the decoherence
functional ensure the
existence of a probability distribution $p_d(\a):=d(\a,\a)$, for all
history propositions $\a$ that belong to the boolean algebra $\cal A$$_{I_n}
$ generated
by the elements $\{\a_i\}$ of $I_n$.\\

The choice of  consistency
conditions adopted here is often referred to as `medium decoherence'. Weaker
conditions require, for example, only the real part of
$d(\a_i,\a_j),i\neq j$, to
vanish; $p_d(\a)$ continues to define a probability function
\cite{Omn92}.
Imposing a weaker consistency
condition implies  that the number of consistent sets for a $d\in\D$
can be larger than the number determined by conditions (\ref{CS}). In
choosing the most restrictive condition we ensure the validity of our
main result for all situations.\\

We concentrate now on the situation where the space $\UP$ is
given by the lattice of projection operators $\PV$ on some
finite-dimensional Hilbert space $\V$.
In this case one can prove the following Classification Theorem for
bounded decoherence functionals:\\

\noindent
{\bf Theorem}\cite{ILS94c} If $\dim\V>2$, decoherence functionals $d$ are
in one-to-one
correspondence with operators $X$ on $\V\otimes\V$ according to the
rule
\beq
        d(\a,\b)=\tr_{\V\otimes\V}(\a\otimes\b X)       \label{GLhist}
\eeq
with the restrictions that:
\beqa
    &a)&\ \tr_{\V\otimes\V}(\a\otimes\b X)=
        \tr_{\V\otimes\V}(\b\otimes\a X^\dagger)
            {\rm\ for\ all\ }\a,\b\in\PV,               \label{herm}\\
    &b)&\ \tr_{\V\otimes\V}(\a\otimes\a X)\ge0
            {\rm\ for\ all\ }\a\in\PV,                  \label{pos} \\
    &c)&\ \tr_{\V\otimes\V}(X)=1.                       \label{norm}
\eeqa

\bigskip
\noindent
This is a non-trivial result since its proof requires Gleason's
Theorem at various stages. Therefore conclusions drawn from it are
themselves non-trivial.

\section{THE MAIN RESULT}
\label{Sec MR}

\subsection{The general case}

Our main result will be valid for any separable Hilbert space despite
the fact that the
classification theorem has not yet been extended to the
infinite-dimensional case. Although it is not clear yet that in this
case every
decoherence functional $d\in\D$ is given by an operator $X$ on $\V\otimes\V$
satisfying the three conditions (\ref{herm} - \ref{norm}), it is
nonetheless true that every operator $X$ fulfilling these conditions
defines a decoherence functional; and this is all that is needed for
the theorem.\\

\noindent
{\bf Theorem}

\medskip
\noindent
Let $\V$ be a separable Hilbert space and let $\UP$ be given by its
lattice of
projection operators $\PV$. Denote by $I=\{\a_1,\a_2,\ldots,\a_n\}$ the set of
projection operators of
an arbitrary partition of unity
and by $p(\a)$ any probability distribution on the boolean algebra $\cal
A$$_{I}$ generated by
this partition. Then there exist bounded decoherence functionals $d_{I}\in\D$
such that $I$ is a consistent set with respect to $d_{I}$ and which
are such that $d_{I}(\a,\a)=p(\a)$ for all $\a\in\cal A$$_{I}$.
\\
Operators $X$ on $\V\otimes\V$ defining such $d_{I}$ are given
by:
\beq
X= K + \sum_{i=1}^{n} p(\a_i)\a_i\otimes\a_i
\eeq
with the following restrictions on the operator $K $:
\beqa
&a)& \tr(\a\otimes\b K)=\tr(\b\otimes\a K^\dagger) \quad\qquad\qquad\!\!\forall
\a,\b \in\PV , \label{M1}\\
&b)& \tr(\b\otimes\b K) + \sum_{i=1}^np(\a_i) [\tr(\b\a_i)]^2 \ge 0\quad\!
\forall \b \notin \{\a_i\}_{i=1}^n, \label{M2}\\
&c)& \tr(\a_i\otimes\a_j K)=0,\quad \tr K=0
\qquad\quad\qquad\!\!\!\!\!\! \forall i, j \in\{1,2,\ldots,n\}.\label{M3}
\eeqa
\bigskip
\noindent
{\bf Proof\/}

\noindent
The proof consists of checking that these operators $X$ have
the required properties. Hence one has to show that conditions
(\ref{herm} - \ref{norm}) are fulfilled to define a decoherence functional.
To prove the consistency conditions (\ref{CS}), one calculates that
$d_X(\a_i,\a_j)=0$. Finally, a short calculation shows that
$d_X(\a_i,\a_i)=p(\a_i)$, which confirms the statement for the
equality of the probability functions.\hfill$\Box$\\

The requirements on $K$ can be met trivially by the zero-operator;
but there exist other solutions. To see that it is not difficult to
write those operators out it is instructive to consider the
finite-dimensional case. This also fixes the notation for the
discussion.

\subsection{The finite-dimensional case}

Let $\V$ be a Hilbert space of dimension $N<\infty$ and let
$\{\ket{e_i}\}_{i=1}^N$ be one of its orthonormal bases. Then a
vector-space basis for the operators on $\V$ is given by
$\{B_{ij}:=\ket{e_i}\bra{e_j}\}_{i,j=1}^N$, so that every operator $A$ on
the tensor product space $\V\otimes\V$ can be expanded as
$A=\sum_{i,j,k,l=1}^N\l_{ij,kl} B_{ij}\otimes B_{kl}$ where
$\l_{ij,kl}\in\mathC$.\\

\noindent
{\bf Lemma}

\medskip
\noindent
Let  dim$\V<\infty$ and let $I_N:=\{\a_1,\a_2,\ldots,\a_N\}$ be a
partition into one-dimensional projectors; denote by $p(\a)$
any probability distribution on the boolean algebra $\cal A$$_{I_N}$
generated by the
elements of $I_N$. Then there exist decoherence functionals
$d_{I_N}\in\D$ such that $I_N$
is a consistent set with respect to these  functionals $d_{I_N}\in\D$
and which are such that  for the probability
distribution $d_{I_N}(\a,\a)=p(\a), \forall\a\in\cal A$$_{I_N}$, holds. \\
Without loss of generality choose the  set $I_N$ to be given by
$I_N:=\{B_{11}, B_{22},\ldots , B_{NN}\}$.
Then the $d_{I_N}$ are given by operators $X$  on
$\V\otimes\V$ of the following form:
\beq
X= K + \sum_{i=1}^N p(B_{ii})B_{ii}\otimes B_{ii} ,\label{x}
\eeq
with the restrictions that:
\beqa
&a)& K=\sum_{i,j=1}^N \k_{ij}B_{ij}\otimes B_{ij},\quad
\k_{ij}\in{\rm I\!R}, \label{hx} \\
&b)& K^T=K, \quad \k_{ii}=0, \quad \k_{ij}\ge 0\quad\forall
i,j\in\{1,2,\ldots,N\}.
\eeqa

\bigskip
\noindent
{\bf Proof\/}

\noindent
It is sufficient to consider only the case when $I_N$ is given by
$I_N:=\{B_{11}, B_{22},\ldots , B_{NN}\}$ because for any other
partition of unity into one-dimensional projectors
$\widetilde{I_N}=\{\widetilde{\a_i}\}_{i=1}^N$ there
exists a unitary transformation $U$ on $\V$ relating both sets via
$UB_{ii}U^\dagger =\widetilde{\a_i}$, for all $i\in\{1,2,\ldots ,N\}$. \\
The corresponding $d_{\widetilde{I_N}}$ are given by the operators $[U\otimes
UXU^\dagger\otimes U^\dagger ]$ on $\V\otimes\V$.\\

In order to show that these $X$ define decoherence
functionals with the required properties, we have to check that  $K$
satisfies the conditions (\ref{M1} -- \ref{M3}).
\be
\item $\tr_{\V\otimes\V}(\a\otimes\b K)= \tr_{\V\otimes\V}(\b\otimes\a
K^\dagger) {\rm\ for\ all\ }\a,\b\in\PV$.
\smallskip\\
The condition $K^T=K$ ensures the symmetry  of $K$.\\
Let $\a=\sum_{i,j=1}^N a_{ij}B_{ij}$ and $\b=\sum_{i,j=1}^N
b_{ij}B_{i j}$ be two arbitrary projectors on $\V$. Using the expression
(\ref{x}) for $X$ one calculates that
\beq
\tr_{\V\otimes\V}(\a\otimes\b K)=\sum_{i,j=1}^N
a_{ji}b_{ji}\k_{ij}=\tr_{\V\otimes\V}(\b\otimes\a K).\label{1}
\eeq
\item $\tr(\b\otimes\b K) + \sum_{i=1}^np(\a_i)[tr(\b\a_i)]^2 \ge 0$ for all
$\b\notin\{\a_i\}_{i=1}^n.$
\smallskip\\
Equation (\ref{1}) shows that $\tr_{\V\otimes\V}(\b\otimes\b K)=\sum_{i,j=1}^N
b_{ji}b_{ji}\k_{ij}$. Since $\b$ is a projection
operator all its
expansion coefficients are real. Therefore $b_{ji}b_{ji}\ge 0, \forall
i,j\in\{1,2,\ldots,N\}$. The condition $\k_{ij}\ge 0$
on the expansion  coefficients of $K$ ensures that
$\tr_{\V\otimes\V}(\b\otimes\b K)\ge0 {\rm\ for\ all\ }\b\in\PV$,
which is sufficient to fulfill the requirement.
\item $\tr_{\V\otimes\V}(K)=0$.
\smallskip\\
This is trivially satisfied because of $\k_{ii}=0$.
\item $\tr_{\V\otimes\V}(B_{ii}\otimes B_{jj}K)=0, \quad\!\! \forall
i,j\in\{1,2,\ldots,N\}$.
\smallskip\\
This condition reflects the consistency conditions
(\ref{CS}), i.e.\  $d_X(B_{ii},B_{jj})=0 \quad\!\!\!\!\!\forall i\neq j$,
and the required equality of the probability functions, i.e.\
$d_X(B_{ii},B_{ii})=p(B_{ii})$.\\
To this end a calculation shows that
\beq
B_{ii}\otimes B_{jj}K=\sum_{\nu,\rho=1}^N \d_{\nu
i}\d_{j\nu}\k_{\nu\rho}B_{i\rho}\otimes B_{j\rho},
\eeq
where $\d_{ij}$ is the Kronecker delta. But there is no way to get rid
of the $\d$'s if $i\neq j$. This means in particular that this
expression is the zero operator whose trace is trivially zero.
In the case $i=j$, one shows that
$\tr_{\V\otimes\V}(B_{ii}\otimes B_{ii}K)=\k_{ii}=0$.
\ee

Therefore the operators $X$ indeed define decoherence functionals via
the rule $d_X(\a,\b)=\tr_{\V\otimes\V}(\a\otimes\b X)$ with the
required properties. \hfill$\Box$\\

Consider now the case when $I_n$ is a partition of unity, which does
{\em not} consist of one-dimensional projectors. Then clearly the
boolean subalgebra $\cal A$$_{I_n}$ with $n\le N$ can be
embedded into a boolean algebra $\cal A$$_{I_N}$ which is generated by
one-dimensional projectors $I_N =\{\a_1,\a_2,\ldots,\a_N\}$. Therefore
this Lemma suffices to prove the following proposition.\\

\noindent
{\bf Proposition}

\medskip
\noindent
Let the dim$\V=N<\infty$. There exists a one-to-many map
$(\cal A $$_{I_n}$,$Prob(\cal A$$_{I_n}$$ )) \mapsto d_{I_N}$ so that
for any pair of a boolean subalgebra
$\cal A$$_{I_n}$ of the space of
history propositions $\UP$ and a probability distribution on this
subalgebra there exist decoherence functionals $d_{I_N}\in\D$, such that
the pairs $(\cal A
$$_{I_n}$,$Prob(\cal A$$_{I_n}$$ ))$ can be obtained from consistent sets
with respect to those $d_{I_N}\in\D$ with $\cal
A$$_{I_n}\subset\cal A$$_{I_N}$.

\section{DISCUSSION}
\label{Sec D}

\subsection{The need for a more refined analysis of the properties of
decoherence functionals}

The theorem established in the last section is mainly an existence
proof. It does {\em not} determine all decoherence functionals
fulfilling the requirements of yielding the given probability function
$p(\a)$ on a given algebra $\cal A$$_{I_N}$. This is because,
at present, no choice of
consistency conditions is known which allows us to find {\em all} boolean
algebras on which $p_d(\a)$ is a probability function. The explicit
construction above of examples of
such $d_{I_N}\in\D$, which have a remarkably simple form,  says that
this subset is already `complete' in the sense that any probability
distribution on the boolean algebra generated by any partition of the
unity in the space of histories $\UP$ can be given by evaluating the
probability function of certain decoherence functionals $d_{I_N}$ on
this algebra.\\

The proposition also does {\em not} show that each decoherence
functional posesses consistent sets. But it has some curious
implications which, for reasons of notational simplicity,
we will discuss only for the finite-dimensional case. \\

Take {\em any} decoherence functional $d\in\D$. Determine its collection of
consistent
sets $\{I_{n_j}^d\}, n\in\{1,2,\ldots,N\}$ and $j\in {\rm I\!N}$, of which
there are $j$ sets with $n$ elements,  with  the
corresponding probability function
$p_d(\a)$. Usually
such a  consistent set will not consist of $N$ one-dimensional
projectors: it will rather be `coarse grained' in the sense that the
dimension of some of the projectors will be greater than one, for
example  for $d_{(H,\rho)}(\a,\b)=\tr_\H(\C_\a^\dagger\rho_{t_0} \C_\b)$.
Embed each boolean algebra $\cal A$$_{I^d_{n_j}}$ generated by one of
these coarse-grained consistent sets into an algebra $\cal A$$_{I_N}$
generated by a partition
of unity into one-dimensional projectors $I_N$. This can
always be done in
{\em many} different ways. Then the theorem  asserts that we can
find at least one simple decoherence functional $d_{I_N}$ which gives
the {\em same} probability function on the subalgebra of the boolean
algebra generated by $I_N$ which is generated by the
elements of $I_{n_j}^d$. Notice that if the consistent set is coarse
grained, there exist uncountably many $d_{I_N}$ having this property.\\

Therefore we have shown that a single pair of a consistent set and a
probability function is in general not sufficient to
uniquely associate  with it a decoherence functional. We therefore
need additional criteria in order to establish  such a correspondence.

Two options appear at present:
\bi
\item The embeddings just mentioned exist for each single
consistent set, but in general there will not be a single embedding
for all the consistent sets belonging to one decoherence functional.
One could therefore start with a {\em collection} of partitions of
unity in $\UP$ and ask if there is a decoherence functional uniquely
associated with it, such that these sets   are consistent.
\item Try to distinguish decoherence functionals by their `symmetry
groups'. This is a notion that will be developed elsewhere. It could
subsume the work on symmetries, equivalence of consistent sets etc.\
\cite{GH94,HLM94,DK94a,DK94b} to tackle
the mentioned problem.
\ei

\subsection{Implications for the history version of standard quantum
mechanics}
\label{Ssec QM}
An example of an orthoalgebra $\UP$ is provided by standard
quantum mechanics. In this case the set of all
history-propositions $\UP$  can indeed be
identified with the lattice of projectors on a certain Hilbert space
or a subset thereof.
Recall that one associates the homogeneous history $\a=\h{\a}$ with a genuine
projection operator
$\a_{t_1}\otimes\a_{t_2}\otimes\cdots\otimes\a_{t_n}$ on the tensor
product space $\V_n :=\H_{t_1}\otimes\H_{t_2}\otimes\cdots\otimes\H_{t_n}$ of
$n$ copies of
the Hilbert space $\H$ on which the canonical theory is defined. The
final Hilbert space is obtained by forming the infinite-dimensional
tensor product $\V:=\otimes^\Omega_{t\in\mathR}\H_t$ as indicated in
\cite{I94a}. Since this space is itself not separable one has to be
cautious about drawing  immediate conclusions from the result for this
case. However,
there is a sense in which it is valid.\\

In \cite{ILS94c} it had been shown that there exists an operator
$X_{(H,\rho)}$ on
$\V\otimes\V$ so that
\beq
        d_{(H,\rho)}(\a,\b)=
            \tr_\H(\C_\a^\dagger\rho_{t_0}
\C_\b)=\tr_{\V\otimes\V}(\a\otimes\b X_{(H,\rho)}). \label{0}
\eeq
In particular if we fix a
temporal support $t_1<t_2<\cdots<t_n$ we can
give the operator explicitly on $\V_n\otimes\V_n$, where $\V_n$ is
the separable Hilbert space mentioned before. If therefore a
consistent set of homogeneous history propositions can
be found for this temporal support, one can equally well describe it
by many other decoherence functionals on $\V_n\otimes\V_n$ so that
$d(\a,\a)=d_{(H,\rho)}(\a,\a)$, albeit not
ones which are of  the standard form (\ref{0}). \\
Here we see once more the need to consider all decoherence
functionals.\\

Let us use the example above to illustrate the kind of `programme' to be
followed in specifying a decoherence functional.

\begin{enumerate}
\item Firstly, we would like
to find a set of (functional) equations
$\{f^i(X)=0\}_{i=1}^l$ which determine $X$ to be of the form $X_{(H,\rho)}$.
\item Secondly, we then need to find  properties of this $X_{(H,\rho)}$ which
characterise the Hamiltonian $H$ and the initial density operator
$\rho$. It is at this point one expects symmetry properties of $X$ to
play a key role.
\end{enumerate}

This procedure is conceptually very clear. One has to
successively specify  properties of decoherence functionals
in addition to their defining ones. Once this task has been solved,
one can ask for selection mechanisms among the consistent sets of
history propositions
belonging to a particular decoherence functional. \\

As indicated in the previous section one can certainly try to
postulate the consistent sets of history propositions and/or the behavior
of the probability function one would like a decoherence functional
to possess in order to accomplish the task of specifying it uniquely by
these means. Suggestions in this direction  have been made by
Isham and Linden in \cite{IL94b}.\\
Omn\`{e}s mentions in his book \cite{Omn94} a
  result along similar lines of thought. But note that he is
already assuming a `unitary evolution' scenario, thereby excluding all
the `non-standard' decoherence functionals from the very start. But
it is certainly of interest if his statement can, under
appropriate assumptions, be translated into this new mathematical
framework. \\

As a first step, as has been suggested by the
referee,  it would be interesting to see if for each partition
of the unit operator into {\em homogeneous} history propositions,
there exists always a {\em standard} decoherence functional,
given by some $X_{(H,\rho)}$, such that this set is a consistent one with a
given probability function. This is a topic for future research.\\

\section{CONCLUSION     }
\label{Sec C}

In assessing the value of the decoherent history programme with respect
to its
potentiality to resolve issues mentioned in the introduction it is
undoubtably necessary to submit its two basic ingredients, the space
of histories $\UP$ and the space of decoherence functionals $\D$, to a
thorough investigation.
The work on the properties of $\UP$ was started in references
\cite{I94a,IL94b}. A systematic approach to the properties of
the space
$\D$ had been initiated in \cite{ILS94c} by the proof of a classification
theorem for decoherence functionals under well defined
circumstances.\\

The present article employs this result to show a completeness
property of the space of decoherence functionals: there are `enough'
decoherence functionals to give each pair
consisting of a partition of unity in $\UP$ and a probability function
on the corresponding algebra the status of a consistent set.\\

Its importance lies in the fact that it calls for an additional
concept, supplementary to the one of `consistency' or `decoherence',
to establish a correspondence between physical systems
and decoherence functionals which is not so general as to be void of
any content. We need to know  more about the properties of decoherence
functionals. \\

It is likely that these new concepts involve notions like `symmetry
groups of decoherence functionals'. To be able to give a sensible
meaning to those expressions one has to develop an analogue of Dirac's
transformation theory for History Theories, a task currently
under investigation.

\bigskip
\bigskip
\noindent
{\large\bf Acknowledgements}

\noindent
It is a pleasure to thank Prof.~C.J.Isham for useful discussions. I
also acknowledge support through a DAAD
fellowship HSPII financed by the German Federal Ministry of Research
and Technology.

\end{document}